\def\aap{{\em Astr.~Astrophys.}}

\def\apj{{\em Astrophys.~J.}}
\def\apjl{{\em Astrophys.~J.~Lett.}}
\def\apjs{{\em Astrophys.~J.~Suppl.}}
\def\apss{{\em Astrophys.~Space~Sci.}}
\def\araa{{\em Ann.~Rev.~Astr.~Astrophys.}}

\def\mnras{{\em Mon.~Not.~R.~astr.~Soc.}}
\def\nat{{\em Nature}}

\def\physrep{{\em Physics Reports}}

\def\procspie{{\em Proc. of the SPIE}}

\def\prd{{\em Phys.~Rev.~D}1}

\def\ssr{{\em Space Sci.~Rev.}}

\def\spose#1{\hbox to 0pt{#1\hss}}
\def\approxlt{\mathrel{\spose{\lower 3pt\hbox{$\sim$}}
        \raise 2.0pt\hbox{$<$}}}
\def\approxgt{\mathrel{\spose{\lower 3pt\hbox{$\sim$}}
        \raise 2.0pt\hbox{$>$}}}

\def\multleft#1{\hbox to size{\vbox {\halign {\lft{##}\cr #1}}\hfill}\par}
\def\multright#1{\hbox to size{\vbox {\halign {\rt{##}\cr #1}}\hfill}\par}

\def\boxit#1{\vbox{\hrule\hbox{\vrule\kern3pt\vbox{\kern3pt
          #1 \kern3pt}\kern3pt\vrule}\hrule}}

\def\K{{\rm\thinspace K}}

\documentclass[11pt,a4]{article}

\usepackage{graphicx}
\usepackage{amsmath}
\usepackage{color}

\date{}

\begin{document}

\title{\bf Observing Black Holes Spin}
\author{Christopher S. Reynolds\\ Institute of Astronomy, Cambridge, CB3 OHA}

\maketitle

\begin{abstract}
  \noindent The spin of a black hole retains the memory of how the
  black hole grew, and can be a potent source of energy for powering
  relativistic jets. To understand the diagnostic power and
  astrophysical significance of black hole spin, however, we must
  first devise observational methods for measuring spin.  Here, I
  describe the current state of black hole spin measurements,
  highlighting the progress made by X-ray astronomers, as
  well as the current excitement of gravitational wave and radio
  astronomy based techniques. Today's spin measurements are already
  constraining models for the growth of supermassive black holes and
  giving new insights into the dynamics of stellar core-collapse,
  {as well as hinting at the physics of relativistic jet
    production.} Future X-ray, radio, and gravitational wave
  observatories will transform black hole spin into a precision tool
  for astrophysics and test fundamental theories of gravity.

\end{abstract}

\noindent A black hole is defined by its event horizon, the point of
no return inside of which even light is destined to be pulled inwards.
Einstein's Theory of General Relativity (GR) tells us that the region
interior to the event horizon follows the Las Vegas rule; what happens
in the black hole interior stays in the black hole interior.  This
causal disconnection has a remarkable consequence; outside of the
event horizon, black holes are nature's simplest objects, defined
solely by their electrical charge (which is neutralized to zero in
realistic astrophysical settings), mass, and angular momentum
\cite{israel:67a, carter:71a}.

The angular momentum, or spin, of a black hole has tremendous
astrophysical importance. An enormous amount of energy can be stored
in the spin and tapped to drive astrophysical processes --- this is
now the standard { theoretical model} for how the
spectacular jets in active galactic nuclei (AGN) are driven
\cite{blandford:77a}, with the spin of a central supermassive black
hole (SMBH) able to energize powerful jets for hundred of millions of
years before becoming exhausted.  In many cases, these jets plough
into gas within their host galaxies, depositing their energy and
stifling star formation \cite{fabian:12a}, making black
hole spin integral to the story of galaxy formation. Spin also
provides a fossil record of how the black hole formed. For example,
the formation of SMBHs in the early Universe is still deeply
mysterious, and whether the population of such black holes is rapidly
spinning or not can differentiate scenarios where they form from
coherent disk-accretion or the chaotic merger of smaller black holes
\cite{volonteri:05a}.

For these reasons, there has been significant interest in developing
methods for measuring the spin of astrophysical black holes
\cite{miller:15a}. In this Review, I will survey the current state and
future promise of black hole spin measurements. For much of the past
20 years, quantitative measures of spin have been the domain of X-ray
astronomy, and these techniques continue to be refined as the quality
of the data improves. With the recent advent of gravitational wave
astronomy, we now have a completely new and complementary window on
spinning black holes.  Furthermore, we stand on the threshold of
another major breakthrough, the direct imaging of the shadow of the
event horizon by global mm-band Very Long Baseline Interferometry,
aka, the Event Horizon Telescope (EHT).  We are truly entering a
golden age for the study of black hole physics and black hole spin.

\section{What is black hole spin?}

Before embarking on a discussion of spin measurements, we must address
a fundamental question --- given the perfect nature of a black hole
(i.e. the complete absence of ``surface'' features), what exactly does
it mean for a black hole to rotate? 

{Assuming that gravity is described by Einstein's General
  Theory of Relativity (GR), }the mathematical description of an
isolated and uncharged spinning black hole was found by Roy Kerr in
the early 1960s \cite{kerr:63a}. The Kerr solution shows that the
spacetime outside of a spinning black hole rotates around the black
hole like a vortex. This rotating spacetime will tend to drag
{nearby} matter, and even light, into rotation around the
black hole. This effect, known as frame-dragging, is weak far from the
black hole and matter is able to resist (if, for example, it has a
rocket motor attached!). Sufficiently close to the spinning black
hole, within a location known as the static limit, the frame-dragging
becomes irresistible and all matter and light is forced to rotate
around the black hole.

This zone of irresistible rotation is important for understanding the
energetics of a spinning black hole.  In a classic thought experiment,
Roger Penrose \cite{penrose:71a} showed that the energy of rotation
resides within this region and can in principle be extracted. This
gives the name to this region --- the ergosphere (derived from the
Ancient Greek {\it ergon} meaning ``work''). While the original
Penrose process may be hard to realise in nature, Roger Blandford and
Roman Znajek showed that magnetic fields can similarly extract
rotational energy from the ergosphere \cite{blandford:77a}. Magnetic
spin-extraction is the leading {theoretical} model for the
driving of relativistic jets from black hole systems.

To be more quantitative, we consider a black hole with mass $M$ and
angular momentum $J$.  We can define the unitless ``spin parameter''
by $a=cJ/GM^2$ where $c$ is the speed of light and $G$ is Newton's
constant of Gravitation. The Kerr solution tells us that the structure
of the spacetime around a spinning black hole depends only on $M$ and
$a$.  {As well as greatly simplifying any GR treatments of
  black hole astrophysics, this provides a route to observational
  explorations of gravity theories beyond GR --- once the mass and
  spin of an astrophysical black hole has been measured, we can in
  principle search for deviations of the inferred gravitational field
  (including any gravitational radiation) from the predictions of GR.}

If one were to spin a planet or a star too quickly, it would fly apart
as the centrifugal forces overwhelm the gravity that binds the object
together.  There is an equivalent situation for a black hole.  The
Kerr solution shows that, if $|a|>1$, there is no longer an event
horizon. GR would then predict a naked spacetime singularity, an
outcome that is abhorrant to physical law and the notion of
predictability and thus forbidden by the Cosmic Censorship
Hypothesis. Of course, it is of great interest to physicists to test
whether nature respects this Kerr limit \cite{bambi:11a}.

\section{Accretion disks, X-rays and black hole spin}

Until recently, spin measurements were only possible for the accreting
SMBHs found in AGN or the accreting stellar mass black holes in X-ray
binaries. In both cases, the accreting gas forms a rotating flattened
disk --- an accretion disk --- whose structure is affected by the
frame-dragging associated with the black hole's spin. The key
challenge is to understand the influence of frame-dragging on
accretion disk structure {and compare these expectations}
to data from real black holes.

\begin{figure}[t]
\begin{center}
\includegraphics[width=0.6\textwidth]{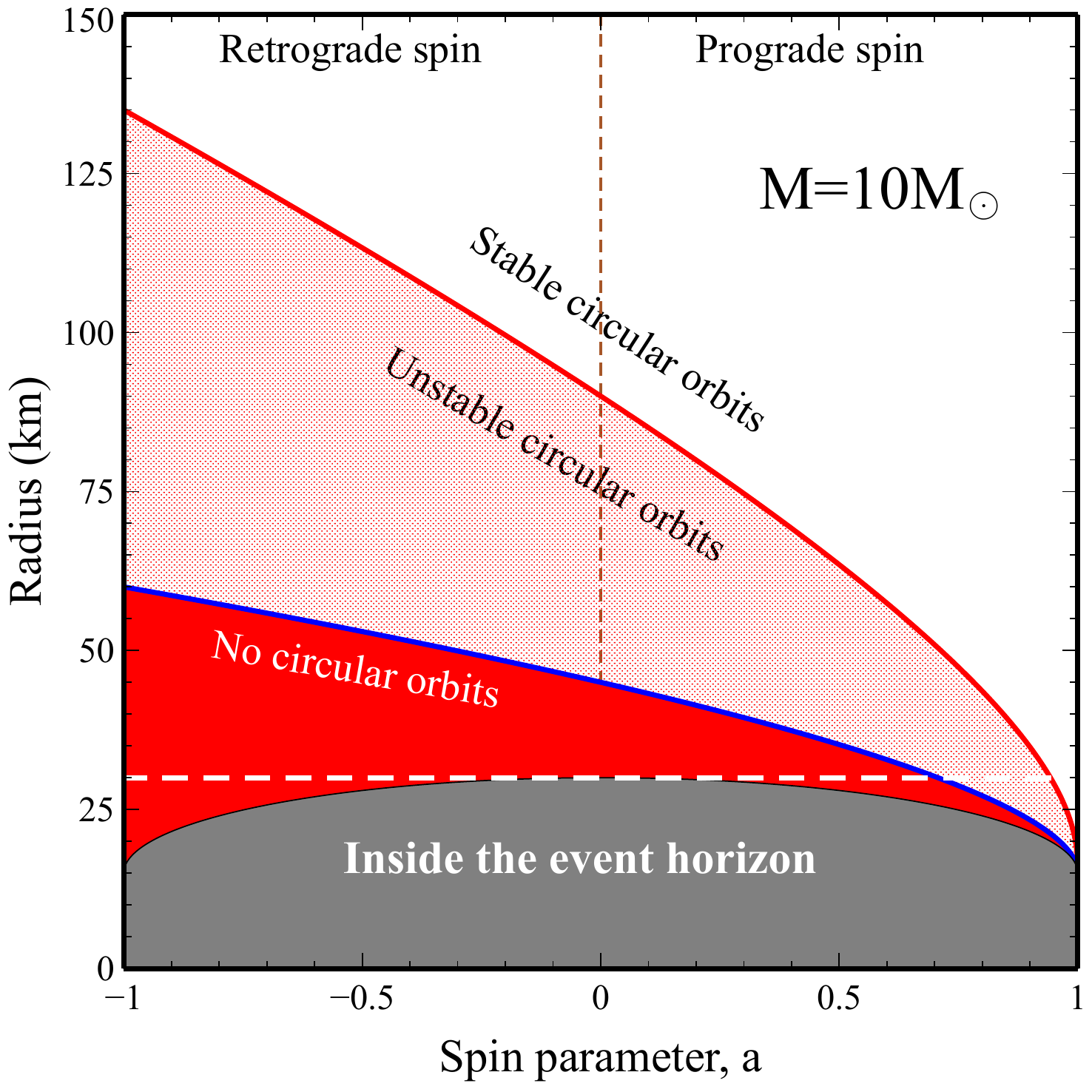}
\end{center}
\caption{{\bf Location of some special orbits in the equatorial plane of a Kerr black hole as a function of spin parameter.}.  Shown here is the innermost stable circular orbit (red line), photon circular orbit (blue line), static limit (dashed white line), and event horizon (bounding the grey shade). Positive/negative spin parameter corresponds to spin that is prograde/retrograde, respectively, relative to the orbiting matter (or photons).  The vertical dashed red line separates the prograde and retrograde cases.   Circular orbits are stable outside of the innermost stable orbit but become unstable inside of this radius (region denoted by light red shading). Circular orbits do not exist interior to the photon circular orbit (region denoted by solid red shading). For concreteness, a 10 solar mass black hole is assumed. Radii for other masses can be obtained using linear proportionality. Figure follows \cite{bardeen:72a}.}
\label{fig:BHradii}
\end{figure}

In general, accretion disks are complex systems and we have yet to
fully understand the influence of black hole spin on their
structure. However, if the accretion rate is in a goldilocks zone ---
not too high and not too low --- the accreting gas can efficiently
radiate the liberated energy as it flows inwards leading to a
geometrically-thin, pancake-like disk \cite{shakura:73a}. Far from the
black hole, where the gravitational field is well approximated by
Newton's law of gravity, the gas in such disks follows stable circular
orbits, with a small inward drift producing the actual accretion.  As
the gas nears the black hole, relativistic effects become important,
de-stabilising circular orbits and causing the accreting gas to
undergo a plunging spiral into the black hole event horizon
\cite{reynolds:08a,shafee:08b}. The location of this transition, the
innermost stable circular orbit (ISCO), is a basic property of the
black hole's gravitational field and depends on the spin
\cite{bardeen:72a}. If the black hole and the accretion disk are
rotating in the same sense, frame-dragging effects can stabilize
otherwise unstable gas motions, resulting in the ISCO being
progressively closer to the black hole as we consider higher spins
(Figure \ref{fig:BHradii}).

\begin{figure}[t]
\begin{center}
\includegraphics[width=0.8\textwidth]{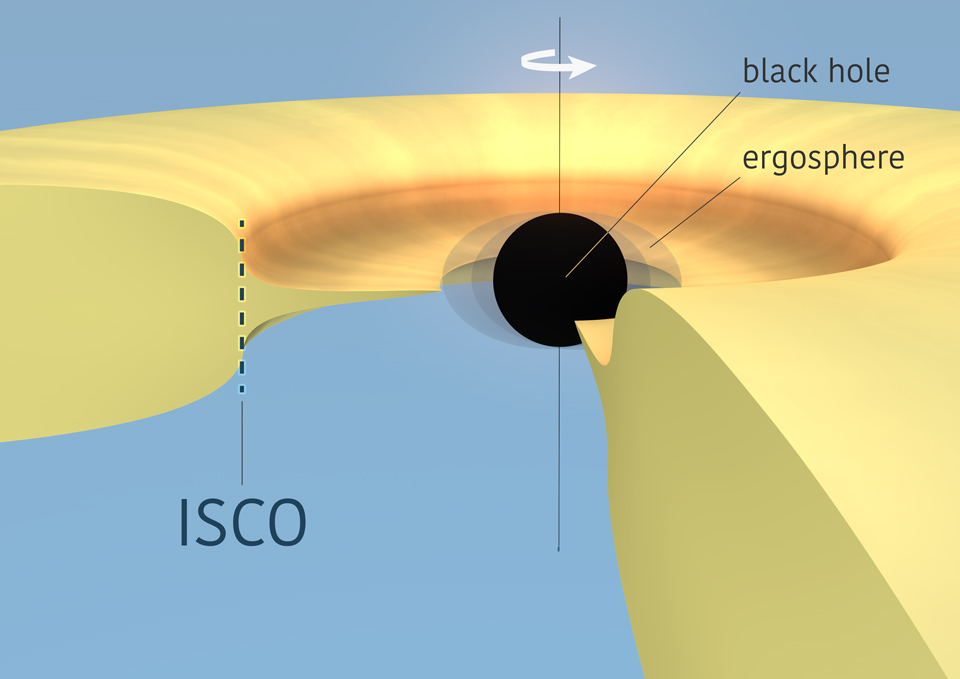}
\end{center}
\caption{{\bf Cartoon of the inner regions of a geometrically-thin
  accretion disk showing the transition in disk structure at the
  innermost stable circular orbit (ISCO).} {Within the ISCO,
    the matter is too highly ionized to produce atomic spectral
    features and the emission processes are non-thermal.} Figure
  courtesy of Amanda Smith (Institute of Astronomy, Cambridge).}
\label{fig:isco_cartoon}
\end{figure}

At least in these geometrically-thin accretion disks, the ISCO
effectively acts as the inner edge to the accretion disk (Figure
\ref{fig:isco_cartoon}) and so imprints itself on the emergent
electromagnetic radiation, especially the X-ray emission from the
inner disk \cite{reynolds:08a,shafee:08b}. {This provides
  the route to measuring spin from X-ray observations.}

The technique employed most extensively to date for measuring spins is
based on gravitational redshifts of atomic features in the X-ray
spectrum. In both AGN and (many) accreting stellar mass black holes in
X-ray binaries, we see powerful high-energy X-ray emissions from a
corona of extremely hot ($T\sim 10^9\K$) electron/positrons pairs
above the accretion disk \cite{fabian:15a}.  The
accretion disk, which is much cooler, is strongly irradiated by these
X-rays causing chemical elements in the surface layers to become
photo-excited and produce fluorescent emission lines at very
well-defined X-ray frequencies \cite{george:91a}; this process is
referred to somewhat inappropriately as X-ray reflection. Given that
we are always viewing the accretion disk at some angle away from
face-on, the observed emission lines from the disk are shifted in
energy by (i) the normal Doppler effect (as the material on its orbit
approaches us and then recedes again leading to blueshifts and
redshifts respectively), (ii) the time-dilation of Special Relativity
(moving clocks run slowly leading to redshifting), and (iii) the
gravitational redshifting of General Relativity \cite{fabian:89a}. The
emission lines become highly broadened and asymmetric, with prominent
blueshifted peaks and long redshifted tails. {The spin of
  the black hole is encoded in these line profiles --- as one
  considers black holes of progressive higher spin, the ISCO moves
  closer to the event horizon, the gravitational redshift of X-rays
  reflected from the ISCO increases, and the extend of the redshifted
  tail of the iron line grows.  \cite{laor:91a,brenneman:06a}. These
  are subtle signatures \cite{reynolds:12a}. X-ray reflection from
  more distant gaseous structures as well as X-ray
  absorption from material in the Milky Way, the host galaxy of the
  AGN, and any accretion disk wind must be carefully modeled and
  accounted for before the subtle effects of spin can be teased from
  the data. In particular, the degeneracy of spin signatures with
  those of accretion disk winds has sparked lively debate
  \cite{miller:09b,reynolds:12b,risaliti:13a}}. Recent work has also explored systematic errors in spin measurements that result from the finite thickness of the accretion disk \cite{taylor:18a}, finding them to become increasingly relevant as one considers black holes with higher accretion rates.

\begin{figure}[t]
\begin{center}
\includegraphics[width=0.8\textwidth]{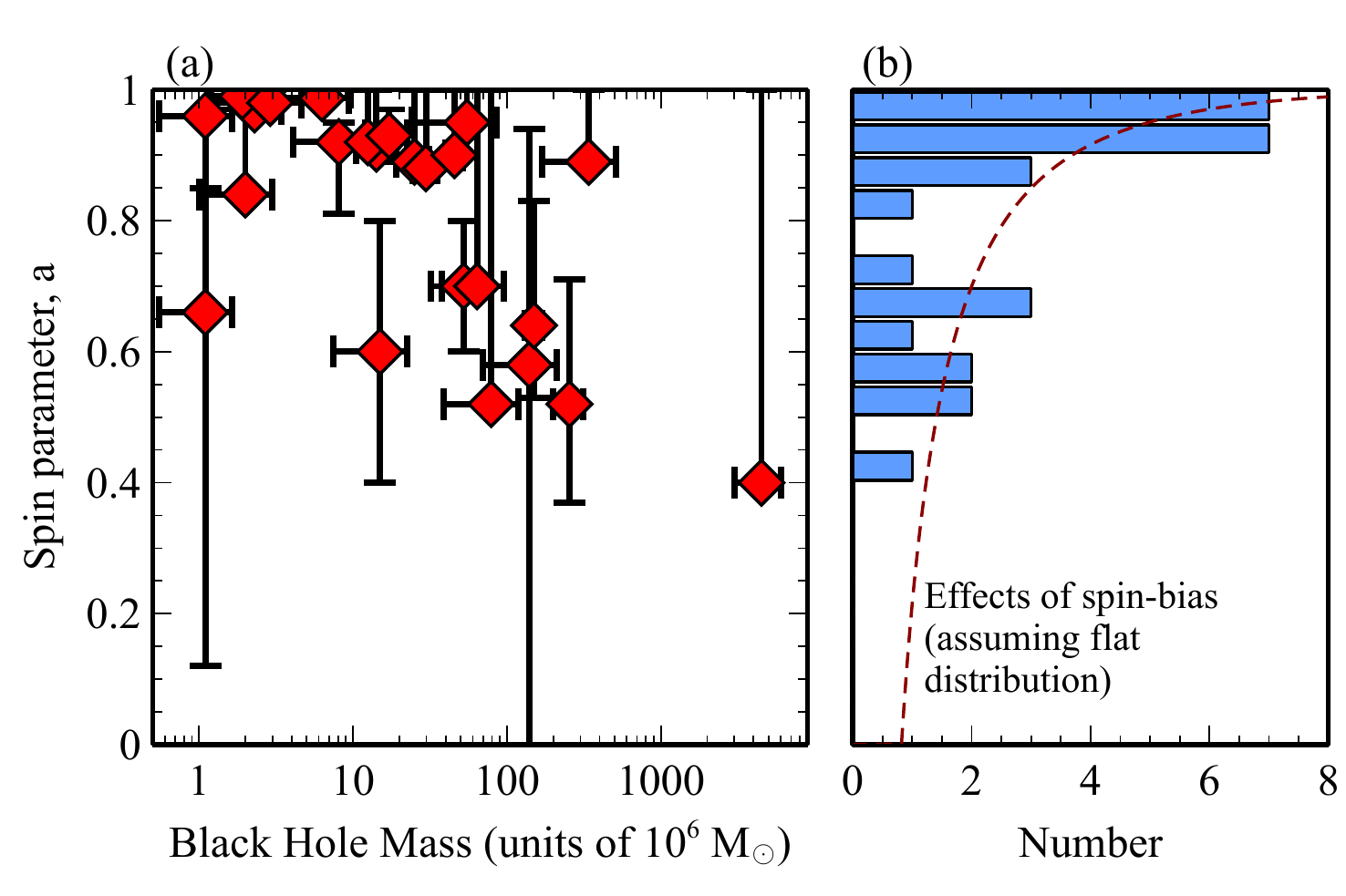}
\end{center}
\caption{{\bf Results on the spins of supermassive black holes in active
  galactic nuclei using the X-ray reflection method.}  Panel-a shows
  the spins and masses of 22 black holes with robust results. Following the conventions in the relevant primary literature, we show 90\% confidence error bars on black hole spin but 1$\sigma$ errors on black hole mass. Panel-b shows the distribution of these spins along with the
  expectations based on a simple efficiency-based selection-bias
  assuming an underlying population of black holes that has equal
  numbers as a function of spin parameter.  Figure based
  on \cite{vasudevan:16a}, {updated with results from
  \cite{ghosh:18a,walton:18a,sun:18a,xu:17a}. Note that four objects
  (IRAS00521--7054, SwiftJ0501.9--32.39, RBS1124, and 1H~0323$+$342)
  have poor SMBH mass constraints; these have been included in the
  spin distribution (right) but not the mass-spin plane.}}
\label{fig:smbh_spins}
\end{figure}

Broadened X-ray reflection has been observed and probed for spin
signatures in both AGN and X-ray binaries for over 20 years
\cite{fabian:89a,tanaka:95a,miller:02a,reynolds:03a,brenneman:06a,miller:07a,walton:12a,risaliti:13a,xu:17a,ghosh:18a,walton:18a,sun:18a}. In
the realm of AGN, there are robust spins for over two dozen SMBHs
\cite{reynolds:14a,vasudevan:16a}. As shown in
Figure~\ref{fig:smbh_spins}a, we find that below about 30
million solar masses, the vast majority of SMBHs examined seem to be
rapidly spinning ($a\approxgt 0.9$). For more massive black holes,
however, we { appear} to pick up a population of more
slowly spinning objects ($a\sim 0.5-0.7$). As discussed later, this
may hint at different formation modes in the two mass regimes.

Interestingly, there may be selection biases which favour finding
high-spin black holes in AGN.  Fundamentally, AGN are chosen for study
because they are bright (or, more technically, they are usually drawn
from flux-limited samples). For a given mass flow rate onto the black
hole, the accretion disk around a more rapidly spinning hole will
release more energy and hence be over-represented in any sample based
on brightness thresholds \cite{vasudevan:16a,brenneman:11a}.  A full
understanding of these selection biases is still ongoing work but, in
a simple treatment, the observed spin distribution (summed over all
masses) is statistically consistent with an underlying population that
has equal numbers of black holes with spins between $a\approx 0.4$ and
the extreme limit ($a=1$), as shown in Figure~\ref{fig:smbh_spins}b.

X-ray reflection based spin measurements do not require knowledge of
either the mass of the black hole or its distance from us since the
gravitational redshift of the ISCO depends only on the spin
parameter. This means that it can be equally applied to AGN (where the
black hole mass is often poorly determined) and the stellar-mass black
holes in X-ray binaries (where the distance can be highly uncertain
\cite{gandhi:18a}). To date, 15 stellar-mass black holes (14 within
our Galaxy and one in the Large Magellanic Cloud) have reflection
based spin measurements, revealing a wide range of spins from rather
slowly rotating ($a=0.3-0.4$) to very rapidly spinning ($a>0.95$)
objects \cite{miller:09a,reynolds:14a}. This observed diversity makes
sense since we expect the selection biases affecting AGN to be absent
from these stellar-mass results --- we essentially know of and have
studied every stellar-mass black hole in our Galaxy that has undergone
an outburst in the past 25 years, so we have a ``fair-sample'' of these objects.

The second major technique that has been extensively used to measure
black hole spin employs the temperature of the accretion disk to
diagnose the location of the ISCO. Specialising again to
geometrically-thin accretion disks, we expect much of the liberated
energy to be radiated as thermal emission (with an approximate
blackbody form) from the disk surface, a disk around a rapidly
spinning black hole possessing higher temperatures (and brighter
emission) than a similar disk around a non-spinning black hole
\cite{novikov:73a,zhang:97a}.  Building models of the
thermal spectrum that take into account the detailed structure of the
accretion disk as well as the influence of the Doppler effect and
gravitational redshifting, we can compare to observations and
determine spin.  Unlike in the gravitational-redshift based methods,
we must also have knowledge of the black hole mass, inclination of the
accretion disk, and distance from the Earth in order to build these
models.

This ``thermal continuum fitting'' technique is particularly effective
for accreting stellar-mass black holes in X-ray binary star
systems. In these systems, the inner accretion disks are hot ($T\sim
10^7\,K$), placing the bulk of their emission in the low-energy X-ray
part of the spectrum where it can penetrate the dust and gas of our
Galaxy, and where we have sensitive detectors on space-based
observatories.  The fact that the gas in the surface layers of these
hot accretion disks is highly ionized leads to only small distortions
from a blackbody spectrum, making the spectral models easier to
calculate and more robust \cite{davis:06a}.  We can also use classical
techniques {in optical astronomy} to extract the black hole
mass and accretion disk inclination from analyses of the orbit of the
companion star \cite{orosz:02a}. Finally, stellar-mass black holes
cycle through different states and so we can wait for them to enter a
state where the thermal emission is completely dominant over the
high-energy X-ray emitting corona \cite{remillard:06a},
validating the assumptions underlying the continuum fitting technique.

To date, 10 stellar-mass black holes (7 in our Galaxy, two in the
Large Magellanic Cloud, and one in the dwarf galaxy M33) have thermal
continuum fitting based spin measurements
\cite{shafee:06a,mcclintock:06a,gou:11a,mcclintock:14a}.
Six of these objects have both thermal continuum and X-ray reflection
based spin measures, allowing us to examine consistency of
results. Only for one object, 4U1543-47, does a significant
discrepancy exist with $a=0.3\pm 0.1$ from X-ray reflection
\cite{miller:09a} and $a=0.8\pm 0.1$ from thermal continuum fitting
\cite{shafee:06a}.  However, both of these measurements were made with
old data from the {\it Rossi X-ray Timing Explorer (RXTE)} which had
limited low-energy X-ray sensitivity and limited spectral
resolution. So it is not surprising if one or both of these
measurements is compromised by systematic errors. {With new
  soft X-ray data coming from the {\it Neutron star Interior
    Composition ExploreR (NICER)} and greatly improved distance
  estimates from {\it Gaia} \cite{gandhi:18a}, we expect a new surge
  of spin results based on continuum-fitting over the next few years.}

Application of the thermal continuum fitting method to AGN is more
challenging. The black hole masses are often highly uncertain, and the
inner accretion disks are relatively cool ($T\sim 10^5\K$) with the
bulk of the thermal emission in the far-or-extreme ultraviolet region
of the spectrum.  This part of the spectrum is strongly absorbed by
the dust and gas in our Galaxy, making it hard to even see this
radiation from an AGN.  Furthermore, atomic processes in the surface
layers of these cool disks also distorts the thermal spectrum away
from blackbody form, making it more challening to calculate robust
models \cite{hubeny:01a}.  Indeed, the surprising degree of optical/UV
variability displayed by AGN has called into question whether a
thermally-emitting standard accretion disk is even the right basic
description for the optical/UV part of the spectrum
\cite{lawrence:18a}.  Still, there have been several studies of SMBH
spin using variants of thermal continuum fitting
\cite{czerny:11a,done:13a,capellupo:17a,piotrovich:17a}. Encouragingly,
for two objects that have well constrained masses as well as both
X-ray reflection and thermal continuum spin measurements, the results
are in good agreement \cite{capellupo:17a}.

\section{Gravitational wave astronomy and black hole spin}

On the 15th September 2015, the advanced Laser Interferometer
Gravitational-Wave Observatory (aLIGO) detected the gravitational
waves as two orbiting 30 solar mass black holes underwent their final
merger in a galaxy more than a billion light years away
\cite{abbott:16a} --- this initiated the era of gravitational wave
astronomy and has opened a new window on black hole spin.

There are three distinct stages to the merger of a binary black hole
system. During the {\it in-spiral} stage, the two black holes orbit
around their common centre of mass, slowly losing energy and angular
momentum via gravitational waves.  The black holes speed up as the
orbit decays leading to a gravitational wave train that becomes
stronger and higher frequency as time goes on.  Eventually, the black
holes get close enough that they pass the binary equivalent of the
ISCO.  They plunge together in a {\it merger} stage that produces
extremely powerful and complex gravitational wave signals. The result
is a single spinning black hole that starts off highly perturbed but
quickly settles down to a Kerr black hole by emitting gravitational
waves in a stage called {\it ring-down}. In principle, the full
gravitational wave signal contains information about the spins of the
two initial black holes, the orientation of these spins relative to
each other and the orbital axis of the binary, and the spin of the
final black hole.  This is also a ``clean'' technique in the sense
that it is pure gravitational evolution and the
inspiral-merger-ringdown wave forms do not depend upon the complex
physics of accretion.

There are of course practical challenges to accessing this clean
information due to the finite signal-to-noise and limited frequency
range of real gravitational wave detectors. For the current generation
of gravitational wave observatories, the target black holes are in the
stellar-mass range and the bulk of the signal is obtained during the
inspiral stage. It can be shown using approximate analytic
calculations (so-called post-Newtonian theory) that the phasing of the
gravitational wave signal during inspiral is influenced by the
mass-weighted average of the individual black hole spins projected
onto the axis around which the black holes are orbiting --- this is
called the effective spin, denoted $\chi_{\rm eff}$
\cite{ajith:11a}. The components of the individual spins that project
into the plane of the orbit cause precession of that plane. This is
known as the precession spin \cite{schmidt:15a}, denoted $\chi_{\rm p}$ and is also
imprinted onto the gravitational wave form but in more subtle manner.

\begin{figure}[t]
\begin{center}
\includegraphics[width=0.5\textwidth,angle=270]{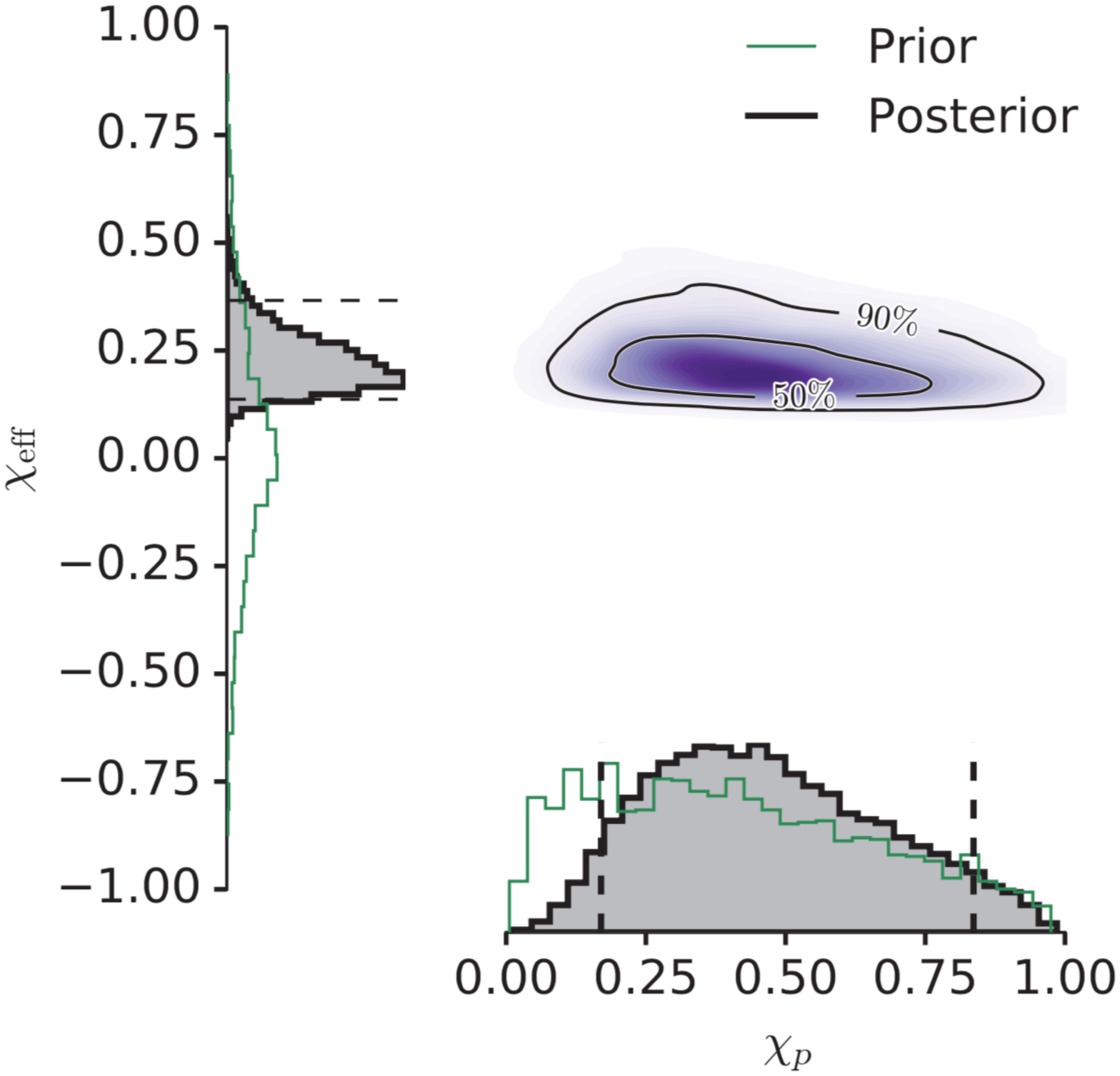}
\end{center}
\caption{{\bf Constraints on black hole spin for GW151226}.  Shown here are the derived probability distribution (black lines with grey shading) and initially assumed prior probability (green) for the effective spin $\chi_{\rm eff}$ and precession spin $\chi_{\rm p}$.  The joint 2-dimensional probability distribution on the $(\chi_{\rm eff}, \chi_{\rm p})$ is shown in purple, along with 50\% and 90\% confidence contours. Figure from \cite{abbott:16b}.}
\label{fig:gw_spin}
\end{figure}

To date, aLIGO working in concert with the European VIRGO
interferometer have high-significance detections of four black hole
binary mergers \cite{abbott:16a,abbott:16b,abbott:17a,abbott:17b}. The
first event GW150915 was consistent with merging non-rotating black
holes \cite{abbott:16c} --- the effective spin was constrained to be
$\chi_{\rm eff}=-0.04^{+0.14}_{-0.16}$ (with 90\% confidence) meaning
that we can rule out rapidly spinning black holes that are both
aligned or anti-aligned with the orbital axis. The constraints on the
precession spin were weak so we cannot tell whether the black holes
are truly rotating slowly, or whether their rapid spins lie in the
plane of the orbit.  In the second event, GW151226, there was a
significant detection of effective spin, showing that at least one of
the black holes has a spin parameter greater $a>0.2$ and that this
spin is at least partially aligned with the orbital axis
\cite{abbott:16b} (Figure~\ref{fig:gw_spin}). Again, though, no
additional information could be gleamed from the precession spin.

We can also examine the spin of the final black hole left once the
merger is complete.  In principle, this spin is imprinted cleanly on
the spectrum of the gravitational waves from ring-down
\cite{detweiler:80a,baibhav:18a}. At current sensitivities, though,
the ring-down emission is difficult to detect and
characterize. However, this final spin is determined to a large degree
by the mass ratio and effective spins of the progenitor binary system.
Taking measurements of these quantities from the inspiral signal,
computer simulations of the full merger process can be used to infer
the final spin typically giving values in the range $a=0.5-0.8$.

\section{The spin of the Galactic Centre black hole}

The closest SMBH to us is in the centre of our own Galaxy, about
25,000 light years away weighing in at 4 million solar masses. To
date, there is no generally accepted measurement of the spin of this
black hole. The accretion rate is extremely low leading to a tenuous
and extremely hot accretion disk that is significantly puffed up into
a fat donut geometry \cite{rees:82a,narayan:95a}, and such an
accretion disk does not produce X-ray reflection features or the
thermal blackbody radiation described above. Also, being a single
black hole, we do not expect any gravitational waves and, even if
there were some unknown companion black hole, the resulting
gravitational waves would be of too low frequency for aLIGO. However,
the unusual proximity of this black hole offers some unique
possibilities for study which are currently being actively developed
\cite{falcke:13a,goddi:17a} and may lead to robust measures of the
spin of this object in the very near future.

Firstly, we have the opportunity to literally take a picture of this
black hole. The event horizon of the Galactic Center black hole will
appear to cast a shadow against the glow of the background accreting
matter \cite{falcke:00a} with an apparent size of 50
micro-arcseconds (14 billionths of a degree, or half the size of a
football on the Moon viewed from the Earth). Such imaging is possible
with the Event Horizon Telescope (EHT), an interferometric experiment
between a network of radio telescopes spread over the Earth operating
at a wavelength of 1.3mm \cite{doeleman:08a,fish:11a}. The fully
operational EHT has conducted observing campaigns on the Galactic
Centre in April 2017 and May 2018 and, at the time of writing, the
data are still being analysed. As well as conclusively proving the
existence of the event horizon for the first time, the EHT images of
the inner accretion disk promise to be a unique probe of the
inclination of the accretion disk, the physical processes that make it
glow, and the black hole spin.  While the size of the shadow is rather
insensitive to the spin, the asymmetry in brightness between the
approaching and receding sides of the disk is expected to be strongly
spin dependent. Indeed, if the black hole has a particularly extreme
spin ($a$ very close to 1), the image of the inner accretion disk
collapses to a vertical line on the approaching side
\cite{gralla:18a}.  However, preliminary data from prototype EHT runs
suggest that this SMBH has a low spin parameter
\cite{broderick:11a,broderick:16a}. There is model dependence to this
result, though, caused by time variability of the accretion disk and
the foreground distribution of Galactic gas that acts as a time
varying scattering screen \cite{medeiros:18a}.  We must await
multi-epoch data from the full EHT before a robust spin can be
derived.

Secondly, the proximity of the Galactic Centre means that we can
observe individual stars orbiting close to the black hole.  As well as
already giving us precision measures of the mass of this black hole
\cite{schodel:02a,ghez:05a,ghez:08a}, it opens the possibility of
seeing the stellar orbits precess due to frame-dragging associated
with the black hole spin. Such measurements require finding and
tracking stars that are within just a few hundred event horizon radii
from the black hole \cite{waisberg:18a}, a task best done with
multiyear near infrared interferometry. This is the main focus of the
GRAVITY instrument \cite{eisenhauer:08a} that has been installed at
the Very Large Telescope Interferometer (VLTI) and operates at
wavelengths of 2--2.4 microns. {The detection of frame-dragging
requires the discovery and subsequent monitoring of stars that are
closer to the SMBH than those currently known \cite{gravity:18a},
currently a difficult task due to the close proximity of the very
bright star known as S2.  Due to the elliptical nature of its orbit,
this star will move away from the SMBH over the course of the next
couple of years thereby allowing us access to these fainter stars that
hold the key to the SMBH's spin.}

\section{Astrophysical implications}

Measurements of black hole spin are providing new insights into the
formation and growth of black holes. As already noted, X-ray
reflection measurements of accreting SMBHs {hint} at two
populations; a rapidly spinning ($a\approxgt 0.9$) population of black
holes that dominate below approximately 30 million solar masses, and a
more massive population of moderately spinning ($a\sim 0.4-0.7$) SMBHs
\cite{reynolds:14a}. {The very existence of a population of
  high-spin SMBHs is interesting, suggesting a significant population
  of SMBHs grew and were spun up by coherent disk accretion. The hint
  that lower-spin objects have higher-mass fits well with the
  hypothesis that today's high-mass SMBHs were built from more
  isotropic chaotic accretion or the merger of smaller black holes
  \cite{volonteri:05a,sesana:14a,fiacconi:18a}.} Implementing these
ideas into future computer models of cosmological galaxy formation
will allow predictions for the spin distribution of the SMBH
population as a function of galaxy mass/type and time since the big
bang.

Moving to the case of stellar-mass black holes, the spin is an
indirect probe of the stellar core-collapse that created the black
hole. The X-ray measurement of moderate- and high-spins in accreting
stellar mass black holes stand in contrast to the very low spin rates
inferred for newly created neutron stars, demonstrating clear
differences in the physics of the supernovae that create these two
classes of objects \cite{miller:11a}. A leading possibility is that
neutron stars experience very strong magnetic braking during or
shortly after their formation; if so, the transfer of energy from the
rotation of the neutron star to the rest of the (exploding) star will
be an important component of the supernovae mechanism.

With even just a small number of detections, gravitational wave
measurements of low effective spin have relevance to the question of
how these binary black holes may form. The most obvious route --- a
binary consisting of two regular massive stars that each evolve and
collapse to black holes --- is problematic.  The tides of one star on
the other would cause them to lock into synchronous rotation (i.e. the
stars would be rotating with the same period as the orbit), resulting
in rapidly spinning black holes aligned with the orbit once the stars
collapse \cite{hotokezaka:17a}. This is incompatible with observed low
effective spins.  Possible solutions invoke binary systems consisting
of unusual Wolf-Rayet stars which are too compact to lock into
synchronous rotation, binary black holes that date back from the very
first generation of star formed in the Universe, {or black
  hole binary formation via dynamical capture processes in dense
  environments such as globular clusters or galactic nuclei
  \cite{miller:02a,stone:17a}.}

Current spin measurements are also providing insights into the physics
of relativistic jet production. Rapidly spinning SMBHs are found in
AGN both with and without powerful jets, proving that black hole spin
is not by itself a sufficient ingredient for jet production (although
it may still be a necessary ingredient). This is not in conflict with
the idea that jets are powered by the magnetic extraction of black
hole spin energy \cite{blandford:77a}, instead suggesting that the
strength and geometry of the magnetic field dictates whether strong
jets are produced.  This magnetic field strength is likely to depend
on the state (temperature and geometric thickness) of the accretion
disk, and hence the accretion rate, raising the possibility that AGN
jets may turn on or off over long timescales. Such a picture is
supported by studies of stellar-mass black holes where it is found
that jets turn on when the accretion rate drops and the accretion disk
becomes tenuous, hot, and geometrically thick.  Turning this around,
it may be possible to estimate spin in strongly jetted sources from
measurement of jet power {if we assume that the
  Blandford-Znajek mechanism is at work.}  Estimates of jet power from
radio observations together with models for the jet launching based on
the Blandford and Znajek mechanism have been used to estimate spin in
powerful radio-galaxies where X-ray reflection signatures are weak and
hard to measure \cite{daly:11a,daly:14a,mikhailov:18a}.

\section{The future}

The coming decades will be tremendously exciting for black hole
astrophysicists. The next generation of X-ray observatories,
especially the future flagship mission {\it ATHENA} (the Advanced
Telescopy for High-ENergy Astrophysics) {that is approved
  and currently under development by the European Space Agency} will
allow further refinement of X-ray reflection and thermal continuum
techniques, providing the sensitivity to study AGN at cosmological
distances when the Universe was appreciably younger. The ability to
measure the spin distribution of the SMBH population as the Universe
ages will strongly constrain our models for how SMBHs first formed and
then grew into today's monsters.

There are also a number of other X-ray techniques that have been
explored for assessing spin in accreting black holes which have
tremendous future potential.  The gravitational microlensing of X-rays
from multiply lensed quasars have already been used for gravitational
redshift-based measures of spin in objects that are too distant and
faint for more traditional X-ray reflection based techniques
\cite{chartas:17a}, and will be a focus of two NASA missions concepts
currently under study, the probe-class Advanced X-ray Imaging
Satellite (AXIS) and the flagship-class LYNX mission. Also, detailed
timing of the variable X-ray signal can in principal measure spin,
both through the detection of X-ray echoes close to the black hole
\cite{reynolds:99a,fabian:09a,zoghbi:12a,kara:13a,cackett:14a} and the
characterisation of the quasi-periodic oscillations sometimes seen
from black holes
\cite{cui:98a,abramowicz:01a,ingram:09a,motta:14a}. Probing
the X-ray echoes will be a major theme of ATHENA and, as well as
giving spin information, allows the structure of the inner accretion
disk to be mapped. The theoretical interpretation of the
quasi-periodic oscillations is still debated, but at least some of
them may be due to spin-induced precession of the inner accretion disk
\cite{fragile:07a,ingram:09a,motta:14a} making them a potent
diagnostic of black hole rotation with ATHENA as well as the
Spectroscopic Time-Resolving Observatory for Broadband Energy X-rays
(STROBE-X) mission {which is currently under study by
  NASA.}

The near future of gravitational wave astronomy will be characterized
by improved sensitivity to merging stellar-mass objects together with
improved ability to localize the source on the sky, enabling follow-up
electromagnetic observations. Longer term, however, we must
dramatically expand the range of frequencies that can be seen. The
current generation of gravitational wave detectors cannot study binary
SMBHs at all due to the restricted range of frequencies over which
they can operate (approximately 10--1000\,Hertz).  The characteristic
frequency of the gravitational waves is inversely proportional to the
mass of the black holes, and so SMBH binaries have frequencies of
milli-Hertz or less, far too low for aLIGO/VIRGO. 

The overall background of very low-frequency gravitational waves
(frequencies of approximately one cycle per year, or 30 nano-Hertz)
from a Universe of SMBH binaries may be first detected through a very
clever route --- the NANOGrav project \cite{arzoumanian:16a} is using
radio telescopes to obtain exquisite timing of a large number of
pulsars (spinning neutron stars) across the sky, seeking correlated
disturbances to their apparent time keeping as a gravitational wave
sweeps over Earth. 

However, the detailed study of gravitational waves from SMBH binaries,
including their spin, must wait for a space-based gravitational wave
observatory such as the Laser Interferometer Space Antenna (LISA)
{which is approved and currently under development by ESA.
  LISA will be} immune from the Earth-bound noise that limits today's
ground-based detectors. Merging SMBHs emit unimaginable amounts of
energy --- during the actual merger event, they are by far the most
energetic process in the entire Universe and can release the binding
energy of an entire galaxy in a few minutes. This power will translate
into high signal-to-noise detections of the gravitational wave train
and precision measures of masses and spins.  We must be aware of
astrophysical selection effects, {though}. SMBHs that find themselves in binary
systems and about to merge have undergone a very special history and
the very process which brings the two black holes together in the
first place may spin them up and/or align them
\cite{bogdanovic:07a}. To gain a full picture of the spins of the SMBH
population, we need both gravitational wave probes of binaries, and
electromagnetic (X-ray) probes of normal accreting SMBHs.

Finally, the ultimate, high-precision measurement of spin will be
possible from the Galactic Centre black hole {\it if only} we
discovered a radio pulsar close to the black hole.  Pulsars are
extremely accurate astrophysical clocks, and the timing of that clock
in orbit around the black hole would allow its orbit to be constructed
with superb accuracy and precision, determining the spin parameter to
0.1\% \cite{liu:12a}. However, we have yet to discovery a suitable
pulsar close to the black hole. We would expect such pulsars to exist,
so the most likely cause for non-detection is the phenomenon of
``dispersion'' by which the observed pulses are smeared out due to
passage through ionized gas in the Galactic Centre.  Dispersion can be
corrected and the pulses de-smeared if we have a very sensitive radio
observatory with excellent frequency resolution.  We expect to detect
a number of such pulsars with the Square Kilometer Array (SKA
\cite{keane:15a}), transforming the precision to which we know the
spin of the Galactic Centre black hole.

\vspace{2cm} 
The author declares no competing financial interests.

Please send correspondence to csr12@ast.cam.ac.uk.


\newpage

\begin{thebibliography}{100}
\expandafter\ifx\csname url\endcsname\relax
  \def\url#1{\texttt{#1}}\fi
\expandafter\ifx\csname urlprefix\endcsname\relax\def\urlprefix{URL }\fi
\providecommand{\bibinfo}[2]{#2}
\providecommand{\eprint}[2][]{\url{#2}}

\bibitem{israel:67a}
\bibinfo{author}{{Israel}, W.}
\newblock \bibinfo{title}{{Event Horizons in Static Vacuum Space-Times}}.
\newblock \emph{\bibinfo{journal}{Physical Review}}
  \textbf{\bibinfo{volume}{164}}, \bibinfo{pages}{1776--1779}
  (\bibinfo{year}{1967}).

\bibitem{carter:71a}
\bibinfo{author}{{Carter}, B.}
\newblock \bibinfo{title}{{Axisymmetric Black Hole Has Only Two Degrees of
  Freedom}}.
\newblock \emph{\bibinfo{journal}{Physical Review Letters}}
  \textbf{\bibinfo{volume}{26}}, \bibinfo{pages}{331--333}
  (\bibinfo{year}{1971}).

\bibitem{blandford:77a}
\bibinfo{author}{{Blandford}, R.~D.} \& \bibinfo{author}{{Znajek}, R.~L.}
\newblock \bibinfo{title}{{Electromagnetic extraction of energy from Kerr black
  holes}}.
\newblock \emph{\bibinfo{journal}{\mnras}} \textbf{\bibinfo{volume}{179}},
  \bibinfo{pages}{433--456} (\bibinfo{year}{1977}).

\bibitem{fabian:12a}
\bibinfo{author}{{Fabian}, A.~C.}
\newblock \bibinfo{title}{{Observational Evidence of Active Galactic Nuclei
  Feedback}}.
\newblock \emph{\bibinfo{journal}{\araa}} \textbf{\bibinfo{volume}{50}},
  \bibinfo{pages}{455--489} (\bibinfo{year}{2012}).

\bibitem{volonteri:05a}
\bibinfo{author}{{Volonteri}, M.}, \bibinfo{author}{{Madau}, P.},
  \bibinfo{author}{{Quataert}, E.} \& \bibinfo{author}{{Rees}, M.~J.}
\newblock \bibinfo{title}{{The Distribution and Cosmic Evolution of Massive
  Black Hole Spins}}.
\newblock \emph{\bibinfo{journal}{\apj}} \textbf{\bibinfo{volume}{620}},
  \bibinfo{pages}{69--77} (\bibinfo{year}{2005}).

\bibitem{miller:15a}
\bibinfo{author}{{Miller}, M.~C.} \& \bibinfo{author}{{Miller}, J.~M.}
\newblock \bibinfo{title}{{The masses and spins of neutron stars and
  stellar-mass black holes}}.
\newblock \emph{\bibinfo{journal}{\physrep}} \textbf{\bibinfo{volume}{548}},
  \bibinfo{pages}{1--34} (\bibinfo{year}{2015}).

\bibitem{kerr:63a}
\bibinfo{author}{{Kerr}, R.~P.}
\newblock \bibinfo{title}{{Gravitational Field of a Spinning Mass as an Example
  of Algebraically Special Metrics}}.
\newblock \emph{\bibinfo{journal}{Physical Review Letters}}
  \textbf{\bibinfo{volume}{11}}, \bibinfo{pages}{237--238}
  (\bibinfo{year}{1963}).

\bibitem{penrose:71a}
\bibinfo{author}{{Penrose}, R.} \& \bibinfo{author}{{Floyd}, R.~M.}
\newblock \bibinfo{title}{{Extraction of Rotational Energy from a Black Hole}}.
\newblock \emph{\bibinfo{journal}{Nature Physical Science}}
  \textbf{\bibinfo{volume}{229}}, \bibinfo{pages}{177--179}
  (\bibinfo{year}{1971}).

\bibitem{bambi:11a}
\bibinfo{author}{{Bambi}, C.}
\newblock \bibinfo{title}{{Spinning super-massive objects in galactic nuclei up
  to a$_{ast}$ $>$ 1}}.
\newblock \emph{\bibinfo{journal}{EPL (Europhysics Letters)}}
  \textbf{\bibinfo{volume}{94}}, \bibinfo{pages}{50002} (\bibinfo{year}{2011}).

\bibitem{bardeen:72a}
\bibinfo{author}{{Bardeen}, J.~M.}, \bibinfo{author}{{Press}, W.~H.} \&
  \bibinfo{author}{{Teukolsky}, S.~A.}
\newblock \bibinfo{title}{{Rotating Black Holes: Locally Nonrotating Frames,
  Energy Extraction, and Scalar Synchrotron Radiation}}.
\newblock \emph{\bibinfo{journal}{\apj}} \textbf{\bibinfo{volume}{178}},
  \bibinfo{pages}{347--370} (\bibinfo{year}{1972}).

\bibitem{shakura:73a}
\bibinfo{author}{{Shakura}, N.~I.} \& \bibinfo{author}{{Sunyaev}, R.~A.}
\newblock \bibinfo{title}{{Black holes in binary systems. Observational
  appearance.}}
\newblock \emph{\bibinfo{journal}{\aap}} \textbf{\bibinfo{volume}{24}},
  \bibinfo{pages}{337--355} (\bibinfo{year}{1973}).

\bibitem{reynolds:08a}
\bibinfo{author}{{Reynolds}, C.~S.} \& \bibinfo{author}{{Fabian}, A.~C.}
\newblock \bibinfo{title}{{Broad Iron-K{$\alpha$} Emission Lines as a
  Diagnostic of Black Hole Spin}}.
\newblock \emph{\bibinfo{journal}{\apj}} \textbf{\bibinfo{volume}{675}},
  \bibinfo{pages}{1048--1056} (\bibinfo{year}{2008}).

\bibitem{shafee:08b}
\bibinfo{author}{{Shafee}, R.} \emph{et~al.}
\newblock \bibinfo{title}{{Three-Dimensional Simulations of Magnetized Thin
  Accretion Disks around Black Holes: Stress in the Plunging Region}}.
\newblock \emph{\bibinfo{journal}{\apjl}} \textbf{\bibinfo{volume}{687}},
  \bibinfo{pages}{L25} (\bibinfo{year}{2008}).

\bibitem{fabian:15a}
\bibinfo{author}{{Fabian}, A.~C.} \emph{et~al.}
\newblock \bibinfo{title}{{Properties of AGN coronae in the NuSTAR era}}.
\newblock \emph{\bibinfo{journal}{\mnras}} \textbf{\bibinfo{volume}{451}},
  \bibinfo{pages}{4375--4383} (\bibinfo{year}{2015}).

\bibitem{george:91a}
\bibinfo{author}{{George}, I.~M.} \& \bibinfo{author}{{Fabian}, A.~C.}
\newblock \bibinfo{title}{{X-ray reflection from cold matter in active galactic
  nuclei and X-ray binaries}}.
\newblock \emph{\bibinfo{journal}{\mnras}} \textbf{\bibinfo{volume}{249}},
  \bibinfo{pages}{352--367} (\bibinfo{year}{1991}).

\bibitem{fabian:89a}
\bibinfo{author}{{Fabian}, A.~C.}, \bibinfo{author}{{Rees}, M.~J.},
  \bibinfo{author}{{Stella}, L.} \& \bibinfo{author}{{White}, N.~E.}
\newblock \bibinfo{title}{{X-ray fluorescence from the inner disc in Cygnus
  X-1}}.
\newblock \emph{\bibinfo{journal}{\mnras}} \textbf{\bibinfo{volume}{238}},
  \bibinfo{pages}{729--736} (\bibinfo{year}{1989}).

\bibitem{laor:91a}
\bibinfo{author}{{Laor}, A.}
\newblock \bibinfo{title}{{Line profiles from a disk around a rotating black
  hole}}.
\newblock \emph{\bibinfo{journal}{\apj}} \textbf{\bibinfo{volume}{376}},
  \bibinfo{pages}{90--94} (\bibinfo{year}{1991}).

\bibitem{brenneman:06a}
\bibinfo{author}{{Brenneman}, L.~W.} \& \bibinfo{author}{{Reynolds}, C.~S.}
\newblock \bibinfo{title}{{Constraining Black Hole Spin via X-Ray
  Spectroscopy}}.
\newblock \emph{\bibinfo{journal}{\apj}} \textbf{\bibinfo{volume}{652}},
  \bibinfo{pages}{1028--1043} (\bibinfo{year}{2006}).

\bibitem{reynolds:12a}
\bibinfo{author}{{Reynolds}, C.~S.} \emph{et~al.}
\newblock \bibinfo{title}{{A Monte Carlo Markov Chain Based Investigation of
  Black Hole Spin in the Active Galaxy NGC 3783}}.
\newblock \emph{\bibinfo{journal}{\apj}} \textbf{\bibinfo{volume}{755}},
  \bibinfo{pages}{88} (\bibinfo{year}{2012}).

\bibitem{miller:09b}
\bibinfo{author}{{Miller}, L.}, \bibinfo{author}{{Turner}, T.~J.} \&
  \bibinfo{author}{{Reeves}, J.~N.}
\newblock \bibinfo{title}{{The absorption-dominated model for the X-ray spectra
  of typeI active galaxies: MCG-6-30-15}}.
\newblock \emph{\bibinfo{journal}{\mnras}} \textbf{\bibinfo{volume}{399}},
  \bibinfo{pages}{L69--L73} (\bibinfo{year}{2009}).

\bibitem{reynolds:12b}
\bibinfo{author}{{Reynolds}, C.~S.}
\newblock \bibinfo{title}{{Constraints on Compton-thick Winds from Black Hole
  Accretion Disks: Can We See the Inner Disk?}}
\newblock \emph{\bibinfo{journal}{\apjl}} \textbf{\bibinfo{volume}{759}},
  \bibinfo{pages}{L15} (\bibinfo{year}{2012}).

\bibitem{risaliti:13a}
\bibinfo{author}{{Risaliti}, G.} \emph{et~al.}
\newblock \bibinfo{title}{{A rapidly spinning supermassive black hole at the
  centre of NGC 1365}}.
\newblock \emph{\bibinfo{journal}{\nat}} \textbf{\bibinfo{volume}{494}},
  \bibinfo{pages}{449--451} (\bibinfo{year}{2013}).

\bibitem{taylor:18a}
\bibinfo{author}{{Taylor}, C.} \& \bibinfo{author}{{Reynolds}, C.~S.}
\newblock \bibinfo{title}{{Exploring the Effects of Disk Thickness on the Black
  Hole Reflection Spectrum}}.
\newblock \emph{\bibinfo{journal}{\apj}} \textbf{\bibinfo{volume}{855}},
  \bibinfo{pages}{120} (\bibinfo{year}{2018}).

\bibitem{tanaka:95a}
\bibinfo{author}{{Tanaka}, Y.} \emph{et~al.}
\newblock \bibinfo{title}{{Gravitationally redshifted emission implying an
  accretion disk and massive black hole in the active galaxy MCG-6-30-15}}.
\newblock \emph{\bibinfo{journal}{\nat}} \textbf{\bibinfo{volume}{375}},
  \bibinfo{pages}{659--661} (\bibinfo{year}{1995}).

\bibitem{miller:02a}
\bibinfo{author}{{Miller}, J.~M.} \emph{et~al.}
\newblock \bibinfo{title}{{Evidence of Spin and Energy Extraction in a Galactic
  Black Hole Candidate: The XMM-Newton/EPIC-pn Spectrum of XTE J1650-500}}.
\newblock \emph{\bibinfo{journal}{\apjl}} \textbf{\bibinfo{volume}{570}},
  \bibinfo{pages}{L69--L73} (\bibinfo{year}{2002}).

\bibitem{reynolds:03a}
\bibinfo{author}{{Reynolds}, C.~S.} \& \bibinfo{author}{{Nowak}, M.~A.}
\newblock \bibinfo{title}{{Fluorescent iron lines as a probe of astrophysical
  black hole systems}}.
\newblock \emph{\bibinfo{journal}{\physrep}} \textbf{\bibinfo{volume}{377}},
  \bibinfo{pages}{389--466} (\bibinfo{year}{2003}).

\bibitem{miller:07a}
\bibinfo{author}{{Miller}, J.~M.}
\newblock \bibinfo{title}{{Relativistic X-Ray Lines from the Inner Accretion
  Disks Around Black Holes}}.
\newblock \emph{\bibinfo{journal}{\araa}} \textbf{\bibinfo{volume}{45}},
  \bibinfo{pages}{441--479} (\bibinfo{year}{2007}).

\bibitem{walton:12a}
\bibinfo{author}{{Walton}, D.~J.}, \bibinfo{author}{{Nardini}, E.},
  \bibinfo{author}{{Fabian}, A.~C.}, \bibinfo{author}{{Gallo}, L.~C.} \&
  \bibinfo{author}{{Reis}, R.~C.}
\newblock \bibinfo{title}{{Suzaku observations of `bare' active galactic
  nuclei}}.
\newblock \emph{\bibinfo{journal}{\mnras}} \textbf{\bibinfo{volume}{428}},
  \bibinfo{pages}{2901--2920} (\bibinfo{year}{2013}).

\bibitem{xu:17a}
\bibinfo{author}{{Xu}, Y.} \emph{et~al.}
\newblock \bibinfo{title}{{Evidence for Relativistic Disk Reflection in the
  Seyfert 1h Galaxy/ULIRG IRAS 05189-2524 Observed by NuSTAR and XMM-Newton}}.
\newblock \emph{\bibinfo{journal}{\apj}} \textbf{\bibinfo{volume}{837}},
  \bibinfo{pages}{21} (\bibinfo{year}{2017}).

\bibitem{ghosh:18a}
\bibinfo{author}{{Ghosh}, R.}, \bibinfo{author}{{Dewangan}, G.~C.},
  \bibinfo{author}{{Mallick}, L.} \& \bibinfo{author}{{Raychaudhuri}, B.}
\newblock \bibinfo{title}{{Broad-band spectral study of the jet-disc emission
  in the radio-loud narrow-line Seyfert 1 galaxy 1H 0323+342}}.
\newblock \emph{\bibinfo{journal}{\mnras}} \textbf{\bibinfo{volume}{479}},
  \bibinfo{pages}{2464--2475} (\bibinfo{year}{2018}).

\bibitem{walton:18a}
\bibinfo{author}{{Walton}, D.~J.} \emph{et~al.}
\newblock \bibinfo{title}{{Disentangling the complex broad-band X-ray spectrum
  of IRAS 13197-1627 with NuSTAR, XMM-Newton and Suzaku}}.
\newblock \emph{\bibinfo{journal}{\mnras}} \textbf{\bibinfo{volume}{473}},
  \bibinfo{pages}{4377--4391} (\bibinfo{year}{2018}).

\bibitem{sun:18a}
\bibinfo{author}{{Sun}, S.} \emph{et~al.}
\newblock \bibinfo{title}{{Multi-epoch analysis of the X-ray spectrum of the
  active galactic nucleus in NGC 5506}}.
\newblock \emph{\bibinfo{journal}{\mnras}} \textbf{\bibinfo{volume}{478}},
  \bibinfo{pages}{1900--1910} (\bibinfo{year}{2018}).

\bibitem{reynolds:14a}
\bibinfo{author}{{Reynolds}, C.~S.}
\newblock \bibinfo{title}{{Measuring Black Hole Spin Using X-Ray Reflection
  Spectroscopy}}.
\newblock \emph{\bibinfo{journal}{\ssr}} \textbf{\bibinfo{volume}{183}},
  \bibinfo{pages}{277--294} (\bibinfo{year}{2014}).

\bibitem{vasudevan:16a}
\bibinfo{author}{{Vasudevan}, R.~V.} \emph{et~al.}
\newblock \bibinfo{title}{{A selection effect boosting the contribution from
  rapidly spinning black holes to the cosmic X-ray background}}.
\newblock \emph{\bibinfo{journal}{\mnras}} \textbf{\bibinfo{volume}{458}},
  \bibinfo{pages}{2012--2023} (\bibinfo{year}{2016}).

\bibitem{brenneman:11a}
\bibinfo{author}{{Brenneman}, L.~W.} \emph{et~al.}
\newblock \bibinfo{title}{{The Spin of the Supermassive Black Hole in NGC
  3783}}.
\newblock \emph{\bibinfo{journal}{\apj}} \textbf{\bibinfo{volume}{736}},
  \bibinfo{pages}{103} (\bibinfo{year}{2011}).

\bibitem{gandhi:18a}
\bibinfo{author}{{Gandhi}, P.} \emph{et~al.}
\newblock \bibinfo{title}{{Gaia DR2 Distances and Peculiar Velocities for
  Galactic Black Hole Transients}}.
\newblock \emph{\bibinfo{journal}{ArXiv e-prints}}  (\bibinfo{year}{2018}).

\bibitem{miller:09a}
\bibinfo{author}{{Miller}, J.~M.}, \bibinfo{author}{{Reynolds}, C.~S.},
  \bibinfo{author}{{Fabian}, A.~C.}, \bibinfo{author}{{Miniutti}, G.} \&
  \bibinfo{author}{{Gallo}, L.~C.}
\newblock \bibinfo{title}{{Stellar-Mass Black Hole Spin Constraints from Disk
  Reflection and Continuum Modeling}}.
\newblock \emph{\bibinfo{journal}{\apj}} \textbf{\bibinfo{volume}{697}},
  \bibinfo{pages}{900--912} (\bibinfo{year}{2009}).

\bibitem{novikov:73a}
\bibinfo{author}{{Novikov}, I.~D.} \& \bibinfo{author}{{Thorne}, K.~S.}
\newblock \bibinfo{title}{{Astrophysics of black holes.}}
\newblock In \bibinfo{editor}{{Dewitt}, C.} \& \bibinfo{editor}{{Dewitt},
  B.~S.} (eds.) \emph{\bibinfo{booktitle}{Black Holes (Les Astres Occlus)}},
  \bibinfo{pages}{343--450} (\bibinfo{year}{1973}).

\bibitem{zhang:97a}
\bibinfo{author}{{Zhang}, S.~N.}, \bibinfo{author}{{Cui}, W.} \&
  \bibinfo{author}{{Chen}, W.}
\newblock \bibinfo{title}{{Black Hole Spin in X-Ray Binaries: Observational
  Consequences}}.
\newblock \emph{\bibinfo{journal}{\apjl}} \textbf{\bibinfo{volume}{482}},
  \bibinfo{pages}{L155--L158} (\bibinfo{year}{1997}).

\bibitem{davis:06a}
\bibinfo{author}{{Davis}, S.~W.} \& \bibinfo{author}{{Hubeny}, I.}
\newblock \bibinfo{title}{{A Grid of Relativistic, Non-LTE Accretion Disk
  Models for Spectral Fitting of Black Hole Binaries}}.
\newblock \emph{\bibinfo{journal}{\apjs}} \textbf{\bibinfo{volume}{164}},
  \bibinfo{pages}{530--535} (\bibinfo{year}{2006}).

\bibitem{orosz:02a}
\bibinfo{author}{{Orosz}, J.~A.} \emph{et~al.}
\newblock \bibinfo{title}{{Dynamical Evidence for a Black Hole in the
  Microquasar XTE J1550-564}}.
\newblock \emph{\bibinfo{journal}{\apj}} \textbf{\bibinfo{volume}{568}},
  \bibinfo{pages}{845--861} (\bibinfo{year}{2002}).

\bibitem{remillard:06a}
\bibinfo{author}{{Remillard}, R.~A.} \& \bibinfo{author}{{McClintock}, J.~E.}
\newblock \bibinfo{title}{{X-Ray Properties of Black-Hole Binaries}}.
\newblock \emph{\bibinfo{journal}{\araa}} \textbf{\bibinfo{volume}{44}},
  \bibinfo{pages}{49--92} (\bibinfo{year}{2006}).

\bibitem{shafee:06a}
\bibinfo{author}{{Shafee}, R.} \emph{et~al.}
\newblock \bibinfo{title}{{Estimating the Spin of Stellar-Mass Black Holes by
  Spectral Fitting of the X-Ray Continuum}}.
\newblock \emph{\bibinfo{journal}{\apjl}} \textbf{\bibinfo{volume}{636}},
  \bibinfo{pages}{L113--L116} (\bibinfo{year}{2006}).

\bibitem{mcclintock:06a}
\bibinfo{author}{{McClintock}, J.~E.} \emph{et~al.}
\newblock \bibinfo{title}{{The Spin of the Near-Extreme Kerr Black Hole GRS
  1915+105}}.
\newblock \emph{\bibinfo{journal}{\apj}} \textbf{\bibinfo{volume}{652}},
  \bibinfo{pages}{518--539} (\bibinfo{year}{2006}).

\bibitem{gou:11a}
\bibinfo{author}{{Gou}, L.} \emph{et~al.}
\newblock \bibinfo{title}{{The Extreme Spin of the Black Hole in Cygnus X-1}}.
\newblock \emph{\bibinfo{journal}{\apj}} \textbf{\bibinfo{volume}{742}},
  \bibinfo{pages}{85} (\bibinfo{year}{2011}).

\bibitem{mcclintock:14a}
\bibinfo{author}{{McClintock}, J.~E.}, \bibinfo{author}{{Narayan}, R.} \&
  \bibinfo{author}{{Steiner}, J.~F.}
\newblock \bibinfo{title}{{Black Hole Spin via Continuum Fitting and the Role
  of Spin in Powering Transient Jets}}.
\newblock \emph{\bibinfo{journal}{\ssr}} \textbf{\bibinfo{volume}{183}},
  \bibinfo{pages}{295--322} (\bibinfo{year}{2014}).

\bibitem{hubeny:01a}
\bibinfo{author}{{Hubeny}, I.}, \bibinfo{author}{{Blaes}, O.},
  \bibinfo{author}{{Krolik}, J.~H.} \& \bibinfo{author}{{Agol}, E.}
\newblock \bibinfo{title}{{Non-LTE Models and Theoretical Spectra of Accretion
  Disks in Active Galactic Nuclei. IV. Effects of Compton Scattering and Metal
  Opacities}}.
\newblock \emph{\bibinfo{journal}{\apj}} \textbf{\bibinfo{volume}{559}},
  \bibinfo{pages}{680--702} (\bibinfo{year}{2001}).

\bibitem{lawrence:18a}
\bibinfo{author}{{Lawrence}, A.}
\newblock \bibinfo{title}{{Quasar viscosity crisis}}.
\newblock \emph{\bibinfo{journal}{Nature Astronomy}}
  \textbf{\bibinfo{volume}{2}}, \bibinfo{pages}{102--103}
  (\bibinfo{year}{2018}).

\bibitem{czerny:11a}
\bibinfo{author}{{Czerny}, B.}, \bibinfo{author}{{Hryniewicz}, K.},
  \bibinfo{author}{{Niko{\l}ajuk}, M.} \& \bibinfo{author}{{S{\c a}dowski}, A.}
\newblock \bibinfo{title}{{Constraints on the black hole spin in the quasar
  SDSS J094533.99+100950.1}}.
\newblock \emph{\bibinfo{journal}{\mnras}} \textbf{\bibinfo{volume}{415}},
  \bibinfo{pages}{2942--2952} (\bibinfo{year}{2011}).

\bibitem{done:13a}
\bibinfo{author}{{Done}, C.}, \bibinfo{author}{{Jin}, C.},
  \bibinfo{author}{{Middleton}, M.} \& \bibinfo{author}{{Ward}, M.}
\newblock \bibinfo{title}{{A new way to measure supermassive black hole spin in
  accretion disc-dominated active galaxies}}.
\newblock \emph{\bibinfo{journal}{\mnras}} \textbf{\bibinfo{volume}{434}},
  \bibinfo{pages}{1955--1963} (\bibinfo{year}{2013}).

\bibitem{capellupo:17a}
\bibinfo{author}{{Capellupo}, D.~M.}, \bibinfo{author}{{Wafflard-Fernandez},
  G.} \& \bibinfo{author}{{Haggard}, D.}
\newblock \bibinfo{title}{{A Comparison of Two Methods for Estimating Black
  Hole Spin in Active Galactic Nuclei}}.
\newblock \emph{\bibinfo{journal}{\apjl}} \textbf{\bibinfo{volume}{836}},
  \bibinfo{pages}{L8} (\bibinfo{year}{2017}).

\bibitem{piotrovich:17a}
\bibinfo{author}{{Piotrovich}, M.~Y.}, \bibinfo{author}{{Gnedin}, Y.~N.},
  \bibinfo{author}{{Natsvlishvili}, T.~M.} \& \bibinfo{author}{{Buliga}, S.~D.}
\newblock \bibinfo{title}{{Constraints on spin of a supermassive black hole in
  quasars with big blue bump}}.
\newblock \emph{\bibinfo{journal}{\apss}} \textbf{\bibinfo{volume}{362}},
  \bibinfo{pages}{231} (\bibinfo{year}{2017}).

\bibitem{abbott:16a}
\bibinfo{author}{{Abbott}, B.~P.} \emph{et~al.}
\newblock \bibinfo{title}{{Observation of Gravitational Waves from a Binary
  Black Hole Merger}}.
\newblock \emph{\bibinfo{journal}{Physical Review Letters}}
  \textbf{\bibinfo{volume}{116}}, \bibinfo{pages}{061102}
  (\bibinfo{year}{2016}).

\bibitem{ajith:11a}
\bibinfo{author}{{Ajith}, P.} \emph{et~al.}
\newblock \bibinfo{title}{{Inspiral-Merger-Ringdown Waveforms for Black-Hole
  Binaries with Nonprecessing Spins}}.
\newblock \emph{\bibinfo{journal}{Physical Review Letters}}
  \textbf{\bibinfo{volume}{106}}, \bibinfo{pages}{241101}
  (\bibinfo{year}{2011}).

\bibitem{schmidt:15a}
\bibinfo{author}{{Schmidt}, P.}, \bibinfo{author}{{Ohme}, F.} \&
  \bibinfo{author}{{Hannam}, M.}
\newblock \bibinfo{title}{{Towards models of gravitational waveforms from
  generic binaries: II. Modelling precession effects with a single effective
  precession parameter}}.
\newblock \emph{\bibinfo{journal}{\prd}} \textbf{\bibinfo{volume}{91}},
  \bibinfo{pages}{024043} (\bibinfo{year}{2015}).

\bibitem{abbott:16b}
\bibinfo{author}{{Abbott}, B.~P.} \emph{et~al.}
\newblock \bibinfo{title}{{GW151226: Observation of Gravitational Waves from a
  22-Solar-Mass Binary Black Hole Coalescence}}.
\newblock \emph{\bibinfo{journal}{Physical Review Letters}}
  \textbf{\bibinfo{volume}{116}}, \bibinfo{pages}{241103}
  (\bibinfo{year}{2016}).

\bibitem{abbott:17a}
\bibinfo{author}{{Abbott}, B.~P.} \emph{et~al.}
\newblock \bibinfo{title}{{GW170104: Observation of a 50-Solar-Mass Binary
  Black Hole Coalescence at Redshift 0.2}}.
\newblock \emph{\bibinfo{journal}{Physical Review Letters}}
  \textbf{\bibinfo{volume}{118}}, \bibinfo{pages}{221101}
  (\bibinfo{year}{2017}).

\bibitem{abbott:17b}
\bibinfo{author}{{Abbott}, B.~P.} \emph{et~al.}
\newblock \bibinfo{title}{{GW170608: Observation of a 19 Solar-mass Binary
  Black Hole Coalescence}}.
\newblock \emph{\bibinfo{journal}{\apjl}} \textbf{\bibinfo{volume}{851}},
  \bibinfo{pages}{L35} (\bibinfo{year}{2017}).

\bibitem{abbott:16c}
\bibinfo{author}{{Abbott}, B.~P.} \emph{et~al.}
\newblock \bibinfo{title}{{Improved Analysis of GW150914 Using a Fully
  Spin-Precessing Waveform Model}}.
\newblock \emph{\bibinfo{journal}{Physical Review X}}
  \textbf{\bibinfo{volume}{6}}, \bibinfo{pages}{041014} (\bibinfo{year}{2016}).

\bibitem{detweiler:80a}
\bibinfo{author}{{Detweiler}, S.}
\newblock \bibinfo{title}{{Black holes and gravitational waves. III - The
  resonant frequencies of rotating holes}}.
\newblock \emph{\bibinfo{journal}{\apj}} \textbf{\bibinfo{volume}{239}},
  \bibinfo{pages}{292--295} (\bibinfo{year}{1980}).

\bibitem{baibhav:18a}
\bibinfo{author}{{Baibhav}, V.}, \bibinfo{author}{{Berti}, E.},
  \bibinfo{author}{{Cardoso}, V.} \& \bibinfo{author}{{Khanna}, G.}
\newblock \bibinfo{title}{{Black hole spectroscopy: Systematic errors and
  ringdown energy estimates}}.
\newblock \emph{\bibinfo{journal}{\prd}} \textbf{\bibinfo{volume}{97}},
  \bibinfo{pages}{044048} (\bibinfo{year}{2018}).

\bibitem{rees:82a}
\bibinfo{author}{{Rees}, M.~J.}, \bibinfo{author}{{Begelman}, M.~C.},
  \bibinfo{author}{{Blandford}, R.~D.} \& \bibinfo{author}{{Phinney}, E.~S.}
\newblock \bibinfo{title}{{Ion-supported tori and the origin of radio jets}}.
\newblock \emph{\bibinfo{journal}{\nat}} \textbf{\bibinfo{volume}{295}},
  \bibinfo{pages}{17--21} (\bibinfo{year}{1982}).

\bibitem{narayan:95a}
\bibinfo{author}{{Narayan}, R.}, \bibinfo{author}{{Yi}, I.} \&
  \bibinfo{author}{{Mahadevan}, R.}
\newblock \bibinfo{title}{{Explaining the spectrum of Sagittarius A$^{*}$ with
  a model of an accreting black hole}}.
\newblock \emph{\bibinfo{journal}{\nat}} \textbf{\bibinfo{volume}{374}},
  \bibinfo{pages}{623--625} (\bibinfo{year}{1995}).

\bibitem{falcke:13a}
\bibinfo{author}{{Falcke}, H.} \& \bibinfo{author}{{Markoff}, S.~B.}
\newblock \bibinfo{title}{{Toward the event horizon --- the supermassive black
  hole in the Galactic Center}}.
\newblock \emph{\bibinfo{journal}{Classical and Quantum Gravity}}
  \textbf{\bibinfo{volume}{30}}, \bibinfo{pages}{244003}
  (\bibinfo{year}{2013}).

\bibitem{goddi:17a}
\bibinfo{author}{{Goddi}, C.} \emph{et~al.}
\newblock \bibinfo{title}{{BlackHoleCam: Fundamental physics of the galactic
  center}}.
\newblock \emph{\bibinfo{journal}{International Journal of Modern Physics D}}
  \textbf{\bibinfo{volume}{26}}, \bibinfo{pages}{1730001--239}
  (\bibinfo{year}{2017}).

\bibitem{falcke:00a}
\bibinfo{author}{{Falcke}, H.}, \bibinfo{author}{{Melia}, F.} \&
  \bibinfo{author}{{Agol}, E.}
\newblock \bibinfo{title}{{Viewing the Shadow of the Black Hole at the Galactic
  Center}}.
\newblock \emph{\bibinfo{journal}{\apjl}} \textbf{\bibinfo{volume}{528}},
  \bibinfo{pages}{L13--L16} (\bibinfo{year}{2000}).

\bibitem{doeleman:08a}
\bibinfo{author}{{Doeleman}, S.~S.} \emph{et~al.}
\newblock \bibinfo{title}{{Event-horizon-scale structure in the supermassive
  black hole candidate at the Galactic Centre}}.
\newblock \emph{\bibinfo{journal}{\nat}} \textbf{\bibinfo{volume}{455}},
  \bibinfo{pages}{78--80} (\bibinfo{year}{2008}).

\bibitem{fish:11a}
\bibinfo{author}{{Fish}, V.~L.} \emph{et~al.}
\newblock \bibinfo{title}{{1.3 mm Wavelength VLBI of Sagittarius A*: Detection
  of Time-variable Emission on Event Horizon Scales}}.
\newblock \emph{\bibinfo{journal}{\apjl}} \textbf{\bibinfo{volume}{727}},
  \bibinfo{pages}{L36} (\bibinfo{year}{2011}).

\bibitem{gralla:18a}
\bibinfo{author}{{Gralla}, S.~E.}, \bibinfo{author}{{Lupsasca}, A.} \&
  \bibinfo{author}{{Strominger}, A.}
\newblock \bibinfo{title}{{Observational signature of high spin at the Event
  Horizon Telescope}}.
\newblock \emph{\bibinfo{journal}{\mnras}} \textbf{\bibinfo{volume}{475}},
  \bibinfo{pages}{3829--3853} (\bibinfo{year}{2018}).

\bibitem{broderick:11a}
\bibinfo{author}{{Broderick}, A.~E.}, \bibinfo{author}{{Fish}, V.~L.},
  \bibinfo{author}{{Doeleman}, S.~S.} \& \bibinfo{author}{{Loeb}, A.}
\newblock \bibinfo{title}{{Evidence for Low Black Hole Spin and Physically
  Motivated Accretion Models from Millimeter-VLBI Observations of Sagittarius
  A*}}.
\newblock \emph{\bibinfo{journal}{\apj}} \textbf{\bibinfo{volume}{735}},
  \bibinfo{pages}{110} (\bibinfo{year}{2011}).

\bibitem{broderick:16a}
\bibinfo{author}{{Broderick}, A.~E.} \emph{et~al.}
\newblock \bibinfo{title}{{Modeling Seven Years of Event Horizon Telescope
  Observations with Radiatively Inefficient Accretion Flow Models}}.
\newblock \emph{\bibinfo{journal}{\apj}} \textbf{\bibinfo{volume}{820}},
  \bibinfo{pages}{137} (\bibinfo{year}{2016}).

\bibitem{medeiros:18a}
\bibinfo{author}{{Medeiros}, L.} \emph{et~al.}
\newblock \bibinfo{title}{{GRMHD Simulations of Visibility Amplitude
  Variability for Event Horizon Telescope Images of Sgr A*}}.
\newblock \emph{\bibinfo{journal}{\apj}} \textbf{\bibinfo{volume}{856}},
  \bibinfo{pages}{163} (\bibinfo{year}{2018}).

\bibitem{schodel:02a}
\bibinfo{author}{{Sch{\"o}del}, R.} \emph{et~al.}
\newblock \bibinfo{title}{{A star in a 15.2-year orbit around the supermassive
  black hole at the centre of the Milky Way}}.
\newblock \emph{\bibinfo{journal}{\nat}} \textbf{\bibinfo{volume}{419}},
  \bibinfo{pages}{694--696} (\bibinfo{year}{2002}).

\bibitem{ghez:05a}
\bibinfo{author}{{Ghez}, A.~M.} \emph{et~al.}
\newblock \bibinfo{title}{{Stellar Orbits around the Galactic Center Black
  Hole}}.
\newblock \emph{\bibinfo{journal}{\apj}} \textbf{\bibinfo{volume}{620}},
  \bibinfo{pages}{744--757} (\bibinfo{year}{2005}).

\bibitem{ghez:08a}
\bibinfo{author}{{Ghez}, A.~M.} \emph{et~al.}
\newblock \bibinfo{title}{{Measuring Distance and Properties of the Milky Way's
  Central Supermassive Black Hole with Stellar Orbits}}.
\newblock \emph{\bibinfo{journal}{\apj}} \textbf{\bibinfo{volume}{689}},
  \bibinfo{pages}{1044--1062} (\bibinfo{year}{2008}).

\bibitem{waisberg:18a}
\bibinfo{author}{{Waisberg}, I.} \emph{et~al.}
\newblock \bibinfo{title}{{What stellar orbit is needed to measure the spin of
  the Galactic centre black hole from astrometric data?}}
\newblock \emph{\bibinfo{journal}{\mnras}} \textbf{\bibinfo{volume}{476}},
  \bibinfo{pages}{3600--3610} (\bibinfo{year}{2018}).

\bibitem{eisenhauer:08a}
\bibinfo{author}{{Eisenhauer}, F.} \emph{et~al.}
\newblock \bibinfo{title}{{GRAVITY: getting to the event horizon of Sgr A*}}.
\newblock In \emph{\bibinfo{booktitle}{Optical and Infrared Interferometry}},
  vol. \bibinfo{volume}{7013} of \emph{\bibinfo{series}{\procspie}},
  \bibinfo{pages}{70132A} (\bibinfo{year}{2008}).

\bibitem{gravity:18a}
\bibinfo{author}{{Gravity Collaboration}} \emph{et~al.}
\newblock \bibinfo{title}{{Detection of the gravitational redshift in the orbit
  of the star S2 near the Galactic centre massive black hole}}.
\newblock \emph{\bibinfo{journal}{\aap}} \textbf{\bibinfo{volume}{615}},
  \bibinfo{pages}{L15} (\bibinfo{year}{2018}).

\bibitem{sesana:14a}
\bibinfo{author}{{Sesana}, A.}, \bibinfo{author}{{Barausse}, E.},
  \bibinfo{author}{{Dotti}, M.} \& \bibinfo{author}{{Rossi}, E.~M.}
\newblock \bibinfo{title}{{Linking the Spin Evolution of Massive Black Holes to
  Galaxy Kinematics}}.
\newblock \emph{\bibinfo{journal}{\apj}} \textbf{\bibinfo{volume}{794}},
  \bibinfo{pages}{104} (\bibinfo{year}{2014}).

\bibitem{fiacconi:18a}
\bibinfo{author}{{Fiacconi}, D.}, \bibinfo{author}{{Sijacki}, D.} \&
  \bibinfo{author}{{Pringle}, J.~E.}
\newblock \bibinfo{title}{{Galactic nuclei evolution with spinning black holes:
  method and implementation}}.
\newblock \emph{\bibinfo{journal}{\mnras}}  (\bibinfo{year}{2018}).

\bibitem{miller:11a}
\bibinfo{author}{{Miller}, J.~M.}, \bibinfo{author}{{Miller}, M.~C.} \&
  \bibinfo{author}{{Reynolds}, C.~S.}
\newblock \bibinfo{title}{{The Angular Momenta of Neutron Stars and Black Holes
  as a Window on Supernovae}}.
\newblock \emph{\bibinfo{journal}{\apjl}} \textbf{\bibinfo{volume}{731}},
  \bibinfo{pages}{L5} (\bibinfo{year}{2011}).

\bibitem{hotokezaka:17a}
\bibinfo{author}{{Hotokezaka}, K.} \& \bibinfo{author}{{Piran}, T.}
\newblock \bibinfo{title}{{Implications of the Low Binary Black Hole Aligned
  Spins Observed by LIGO}}.
\newblock \emph{\bibinfo{journal}{\apj}} \textbf{\bibinfo{volume}{842}},
  \bibinfo{pages}{111} (\bibinfo{year}{2017}).

\bibitem{stone:17a}
\bibinfo{author}{{Stone}, N.~C.}, \bibinfo{author}{{K{\"u}pper}, A.~H.~W.} \&
  \bibinfo{author}{{Ostriker}, J.~P.}
\newblock \bibinfo{title}{{Formation of massive black holes in galactic nuclei:
  runaway tidal encounters}}.
\newblock \emph{\bibinfo{journal}{\mnras}} \textbf{\bibinfo{volume}{467}},
  \bibinfo{pages}{4180--4199} (\bibinfo{year}{2017}).

\bibitem{daly:11a}
\bibinfo{author}{{Daly}, R.~A.}
\newblock \bibinfo{title}{{Estimates of black hole spin properties of 55
  sources}}.
\newblock \emph{\bibinfo{journal}{\mnras}} \textbf{\bibinfo{volume}{414}},
  \bibinfo{pages}{1253--1262} (\bibinfo{year}{2011}).

\bibitem{daly:14a}
\bibinfo{author}{{Daly}, R.~A.} \& \bibinfo{author}{{Sprinkle}, T.~B.}
\newblock \bibinfo{title}{{Black hole spin properties of 130 AGN}}.
\newblock \emph{\bibinfo{journal}{\mnras}} \textbf{\bibinfo{volume}{438}},
  \bibinfo{pages}{3233--3242} (\bibinfo{year}{2014}).

\bibitem{mikhailov:18a}
\bibinfo{author}{{Mikhailov}, A.~G.} \& \bibinfo{author}{{Gnedin}, Y.~N.}
\newblock \bibinfo{title}{{Determination of the Spins of Supermassive Black
  Holes in FR I and FR II Radio Galaxies}}.
\newblock \emph{\bibinfo{journal}{Astronomy Reports}}
  \textbf{\bibinfo{volume}{62}}, \bibinfo{pages}{1--8} (\bibinfo{year}{2018}).

\bibitem{chartas:17a}
\bibinfo{author}{{Chartas}, G.} \emph{et~al.}
\newblock \bibinfo{title}{{Measuring the Innermost Stable Circular Orbits of
  Supermassive Black Holes}}.
\newblock \emph{\bibinfo{journal}{\apj}} \textbf{\bibinfo{volume}{837}},
  \bibinfo{pages}{26} (\bibinfo{year}{2017}).

\bibitem{reynolds:99a}
\bibinfo{author}{{Reynolds}, C.~S.}, \bibinfo{author}{{Young}, A.~J.},
  \bibinfo{author}{{Begelman}, M.~C.} \& \bibinfo{author}{{Fabian}, A.~C.}
\newblock \bibinfo{title}{{X-Ray Iron Line Reverberation from Black Hole
  Accretion Disks}}.
\newblock \emph{\bibinfo{journal}{\apj}} \textbf{\bibinfo{volume}{514}},
  \bibinfo{pages}{164--179} (\bibinfo{year}{1999}).

\bibitem{fabian:09a}
\bibinfo{author}{{Fabian}, A.~C.} \emph{et~al.}
\newblock \bibinfo{title}{{Broad line emission from iron K- and L-shell
  transitions in the active galaxy 1H0707-495}}.
\newblock \emph{\bibinfo{journal}{\nat}} \textbf{\bibinfo{volume}{459}},
  \bibinfo{pages}{540--542} (\bibinfo{year}{2009}).

\bibitem{zoghbi:12a}
\bibinfo{author}{{Zoghbi}, A.}, \bibinfo{author}{{Fabian}, A.~C.},
  \bibinfo{author}{{Reynolds}, C.~S.} \& \bibinfo{author}{{Cackett}, E.~M.}
\newblock \bibinfo{title}{{Relativistic iron K X-ray reverberation in NGC
  4151}}.
\newblock \emph{\bibinfo{journal}{\mnras}} \textbf{\bibinfo{volume}{422}},
  \bibinfo{pages}{129--134} (\bibinfo{year}{2012}).

\bibitem{kara:13a}
\bibinfo{author}{{Kara}, E.} \emph{et~al.}
\newblock \bibinfo{title}{{Discovery of high-frequency iron K lags in Ark 564
  and Mrk 335}}.
\newblock \emph{\bibinfo{journal}{\mnras}} \textbf{\bibinfo{volume}{434}},
  \bibinfo{pages}{1129--1137} (\bibinfo{year}{2013}).

\bibitem{cackett:14a}
\bibinfo{author}{{Cackett}, E.~M.} \emph{et~al.}
\newblock \bibinfo{title}{{Modelling the broad Fe K{$\alpha$} reverberation in
  the AGN NGC 4151}}.
\newblock \emph{\bibinfo{journal}{\mnras}} \textbf{\bibinfo{volume}{438}},
  \bibinfo{pages}{2980--2994} (\bibinfo{year}{2014}).

\bibitem{cui:98a}
\bibinfo{author}{{Cui}, W.}, \bibinfo{author}{{Zhang}, S.~N.} \&
  \bibinfo{author}{{Chen}, W.}
\newblock \bibinfo{title}{{Evidence for Frame Dragging around Spinning Black
  Holes in X-Ray Binaries}}.
\newblock \emph{\bibinfo{journal}{\apjl}} \textbf{\bibinfo{volume}{492}},
  \bibinfo{pages}{L53--L57} (\bibinfo{year}{1998}).

\bibitem{abramowicz:01a}
\bibinfo{author}{{Abramowicz}, M.~A.} \& \bibinfo{author}{{Klu{\'z}niak}, W.}
\newblock \bibinfo{title}{{A precise determination of black hole spin in GRO
  J1655-40}}.
\newblock \emph{\bibinfo{journal}{\aap}} \textbf{\bibinfo{volume}{374}},
  \bibinfo{pages}{L19--L20} (\bibinfo{year}{2001}).

\bibitem{ingram:09a}
\bibinfo{author}{{Ingram}, A.}, \bibinfo{author}{{Done}, C.} \&
  \bibinfo{author}{{Fragile}, P.~C.}
\newblock \bibinfo{title}{{Low-frequency quasi-periodic oscillations spectra
  and Lense-Thirring precession}}.
\newblock \emph{\bibinfo{journal}{\mnras}} \textbf{\bibinfo{volume}{397}},
  \bibinfo{pages}{L101--L105} (\bibinfo{year}{2009}).

\bibitem{motta:14a}
\bibinfo{author}{{Motta}, S.~E.} \emph{et~al.}
\newblock \bibinfo{title}{{Black hole spin measurements through the
  relativistic precession model: XTE J1550-564}}.
\newblock \emph{\bibinfo{journal}{\mnras}} \textbf{\bibinfo{volume}{439}},
  \bibinfo{pages}{L65--L69} (\bibinfo{year}{2014}).

\bibitem{fragile:07a}
\bibinfo{author}{{Fragile}, P.~C.}, \bibinfo{author}{{Blaes}, O.~M.},
  \bibinfo{author}{{Anninos}, P.} \& \bibinfo{author}{{Salmonson}, J.~D.}
\newblock \bibinfo{title}{{Global General Relativistic Magnetohydrodynamic
  Simulation of a Tilted Black Hole Accretion Disk}}.
\newblock \emph{\bibinfo{journal}{\apj}} \textbf{\bibinfo{volume}{668}},
  \bibinfo{pages}{417--429} (\bibinfo{year}{2007}).

\bibitem{arzoumanian:16a}
\bibinfo{author}{{Arzoumanian}, Z.} \emph{et~al.}
\newblock \bibinfo{title}{{The NANOGrav Nine-year Data Set: Limits on the
  Isotropic Stochastic Gravitational Wave Background}}.
\newblock \emph{\bibinfo{journal}{\apj}} \textbf{\bibinfo{volume}{821}},
  \bibinfo{pages}{13} (\bibinfo{year}{2016}).

\bibitem{bogdanovic:07a}
\bibinfo{author}{{Bogdanovi{\'c}}, T.}, \bibinfo{author}{{Reynolds}, C.~S.} \&
  \bibinfo{author}{{Miller}, M.~C.}
\newblock \bibinfo{title}{{Alignment of the Spins of Supermassive Black Holes
  Prior to Coalescence}}.
\newblock \emph{\bibinfo{journal}{\apjl}} \textbf{\bibinfo{volume}{661}},
  \bibinfo{pages}{L147--L150} (\bibinfo{year}{2007}).

\bibitem{liu:12a}
\bibinfo{author}{{Liu}, K.}, \bibinfo{author}{{Wex}, N.},
  \bibinfo{author}{{Kramer}, M.}, \bibinfo{author}{{Cordes}, J.~M.} \&
  \bibinfo{author}{{Lazio}, T.~J.~W.}
\newblock \bibinfo{title}{{Prospects for Probing the Spacetime of Sgr A* with
  Pulsars}}.
\newblock \emph{\bibinfo{journal}{\apj}} \textbf{\bibinfo{volume}{747}},
  \bibinfo{pages}{1} (\bibinfo{year}{2012}).

\bibitem{keane:15a}
\bibinfo{author}{{Keane}, E.} \emph{et~al.}
\newblock \bibinfo{title}{{A Cosmic Census of Radio Pulsars with the SKA}}.
\newblock \emph{\bibinfo{journal}{Advancing Astrophysics with the Square
  Kilometre Array (AASKA14)}} \bibinfo{pages}{40} (\bibinfo{year}{2015}).

\end{thebibliography}
\end{document}